\documentclass[10pt,preprint]{aastex}
\usepackage{float,amsmath}
\usepackage{ulem,color}

\newcommand{\teff}{$T_{\rm eff}$}
\newcommand{\tc}{$T_{\mathrm{C}}$}
\newcommand{\ME}{$M_{\oplus}$}

\newcommand{\hda}{HD\,80606}

\newcommand{\hdb}{HD\,80607}
\newcommand{\hdab}{HD\,80606/07}

\newcommand{\hdatetwo}{HD\,20782}
\newcommand{\hdateone}{HD\,20781}
\newcommand{\hdatetwoone}{HD\,20782/81}

\newcommand{\kgsins}[1]{{#1}}
\newcommand{\kgsinsplain}[1]{{#1}}
\newcommand{\kgsdel}[1]{\iffalse {#1} \fi}

\begin{document}
\title{Detailed Abundances of Planet-Hosting Wide Binaries.\ II.\
HD\,80606+HD\,80607}


\author{
Claude E.\ Mack III\altaffilmark{1}, 
Keivan G.\ Stassun\altaffilmark{1,2},
Simon C.\ Schuler\altaffilmark{3},
Leslie Hebb\altaffilmark{4},
Joshua A.\ Pepper\altaffilmark{5}
}
\affil{
\altaffiltext{1}{Vanderbilt University, Department of Physics \& Astronomy, Nashville, TN 37235 USA}
\altaffiltext{2}{Fisk University, Physics Department, Nashville, TN 37208 USA; keivan.stassun@vanderbilt.edu}
\altaffiltext{3}{University of Tampa, Department of Chemistry, Biochemistry, and Physics, Tampa, FL 33606 USA}
\altaffiltext{4}{Hobart and William Smith Colleges, Department of Physics, Geneva, NY 14456 USA}
\altaffiltext{5}{Lehigh University, Department of Physics, Bethlehem, PA 18015 USA}
}


\begin{abstract}
We present a detailed chemical abundance analysis of 15 elements 
in
the planet-hosting wide binary system \hda\ + \hdb\ using 
Keck/HIRES spectra. 
As in our previous analysis of the planet-hosting wide binary \hdatetwo\ + \hdateone, we presume that these two G5 dwarf stars formed 
together and therefore had identical primordial abundances. 
In this binary, 
\hda\ 
hosts an eccentric ($e\approx0.93$) giant planet at $\sim$0.5 AU, but \hdb\ has no detected planets. If close-in giant planets on eccentric orbits are efficient at scattering rocky planetary material into their host stars, then \hda\ should show evidence of having accreted rocky material while \hdb\ should not. Here we show that the trends of abundance versus element condensation temperature for \hda\ and \hdb\ are statistically indistinguishable, \kgsins{corroborating the recent result of Saffe et al}. This could suggest that both stars accreted similar amounts of rocky material; indeed, our model for the chemical signature of rocky planet accretion 
indicates that \hda\ could have accreted up to 
\kgsins{2.5}~$M_{\oplus}$ of rocky 
material\kgsins{---about half that contained in the Solar System and primordial asteroid belt---}relative to \hdb\ and still be consistent with the data. 
\kgsinsplain{Since \hdb\ has no known giant planets that might have pushed rocky planet material via migration onto that star,}
we consider it more likely that \hdab\ experienced essentially no rocky planet accretion. This in turn suggests that the migration history of the \hda\ giant planet must have been such that it ejected \kgsinsplain{any} close-in planetary material 
\kgsinsplain{that might have otherwise been} shepherded 
onto the star. 
\end{abstract}


\section{Introduction\label{s:intro}}

This paper is the second in a series of papers that aims to investigate the relationship between the architectures of planetary systems and the detailed chemical composition of the host star by directly comparing the chemical abundances of each stellar pair in \kgsins{planet-hosting wide binaries (PHWB)}. In the first paper \citep{mack14}, we analyzed the detailed chemical abundances of the PHWB \hdatetwo\ + \hdateone. For both stars in that system, we found a trend between the elemental abundances ([X/H]) and the elemental condensation temperature (\tc) that, according to a simple model for the \kgsins{accretion of H-depleted rocky planetary material} by a solar-type star, was consistent with the ingestion of 10--20 $M_{\rm \oplus}$ by each star.

{In this paper, we present the analysis of the detailed abundance trends in the two stars comprising the \hdab\ system.} \hda\ and \hdb\ are a common proper motion wide binary with an angular separation of $20.6''$ and a projected physical separation of $\sim1200$ AU ({Raghavan~et~al.~2006}). They are both solar-type stars with the same spectral type, G5V, and {and apparent V magnitudes of 9.06 and 9.17}, respectively~\citep{2001KFNT...17..409K,1997ESASP1200.....E}. \hda\ hosts a $\sim4\,\,M_{\rm Jup}$ on a very eccentric ($e\sim0.93$) orbit at $\sim$0.5 AU~\citep{naef01,2009A&A...502..695P}. On the other {hand}, there are no detected planetary-mass companions to \hdb. 

If the presence of a close-in eccentric giant planet is an indicator that the host star has likely ingested a significant \kgsins{amount of H-depleted material,} as may be the case with \hdatetwoone\ and a subset of single stars analyzed by \citet{schuler11a}, then one might expect that the chemical abundances of \hda\ might show evidence of rocky planetary accretion, while \hdb\ would not. If the chemical abundances of \hdab\ are not found to be distinct as was the case in the recent study performed by \citet{saffe15}, then this suggests that there are other factors besides the present-day architecture of the planetary system that determine the amount of H-depleted rocky material that is ingested by a planet-hosting star.  One of these factors might be the specific migration history of the giant planets in the system. As a result, not all host stars with close-in giant planets would necessarily exhibit the abundance trends seen in \hdatetwoone. Indeed, the degree of the correlation between planetary-system architecture and the chemical composition of the host star is the main goal of this ongoing series of papers. If a correlation is discovered, then while we cannot expect it to be true for each specific system, it could still serve as tool for more targeted searches for Solar System analogs in general.

In Section~\ref{s:data}, we describe our observations and spectral analysis.
In Section~\ref{s:results}, we present the main results, 
In Section~\ref{s:disc} we discuss the results
in the context of a simple model for how the accretion of Earth-like
rocky planets would affect refractory elemental abundances as a function of \tc,
as well as the results of numerical simulations found in the literature, that
investigation how giant planet migration
affects planetary material interior to the orbit of the giant.
Finally, in Section~\ref{s:conc} we briefly summarize the main conclusions.

\section{Data and Analysis\label{s:data}}
{For both \hdab, on UT 2011 Mar 14 we obtained} high-resolution, 
high-signal-to-noise ratio (SNR) spectra with the 10-m 
Keck I telescope and the HIRES echelle spectrograph~\citep{1994SPIE.2198..362V}
{in the $R=\lambda / \Delta \lambda=72,000$ mode}.
We used the kv418 filter with the B2 slit setting ($0.574''$ $\times$ $7''$)
and 2$\times$1 binning (spatial $\times$ dispersion).
{The spectra cover a wavelength range from $\sim3500-9500$\AA.}
For each of the stars in the system, one exposure was taken  
with an integration time of 900s. 
The SNR in the continuum region near $\lambda 6700$ of $\sim$300 for \hdb\ and $\sim$330 for \hda.
{The data were reduced using the {\sc makee} data reduction routines.}
A sample spectrum spanning the wavelength region $\lambda$6135-$\lambda$6175 is
shown in Figure~\ref{fig:spec0607}. \kgsins{We note that these spectra were also used by \citet{saffe15} for their abundance analysis of these stars.}
{We were unable to obtain a solar spectrum during the same observing
run that we observed \hdab.  Therefore, for the derivation of the relative
chemical abundances, we used a previous Keck/HIRES solar
spectrum that we obtained in 2010 June. The solar spectrum
has an SNR of $\sim$800 near $\lambda 6700$.

For both \hdab, chemical abundances relative to solar have been derived from
the observed spectra for 15 elements.
We used the 2010 version of the LTE spectral analysis package {\sc moog} \citep{1973PhDT.......180S}
abundances were derived from measurements of the equivalent widths (EWs) of atomic lines
using the {\sc spectre} analysis package~\citep{1987BAAS...19.1129F}. Stellar parameters 
were obtained by requiring excitation and ionization balance of 
the \ion{Fe}{1} and \ion{Fe}{2} lines.
{The atomic excitation energies ($\chi$) and transition probabilities ($\log{gf}$)
were taken from the Vienna Atomic Line Database~\citep[VALD;][]{1995A&AS..112..525P,1999A&AS..138..119K}}.

For the odd-Z elements V, Mn, and Co, in order to take into account
hyper-fine structure (hfs) effects {\citep{2000ApJ...537L..57P}},
spectral synthesis incorporating hfs components has been used to compare with
the EW-based abundances.  {The hfs components for these elements were obtained
from \citet{2006ApJ...640..801J}, and the line lists for wavelength regions encompassing each
feature were taken from VALD.} The adopted V, Mn, and Co abundances are derived from
the hfs analysis and those lines with EWs that are less affected 
by hfs.    

The abundance and error analyses {for all elements} are described in greater detail in
\citet{schuler11a}.  The analysis we performed was nearly identical to our
analysis of the PHWB HD\,20782/81 \citep{mack14}.
The stellar parameters and relative abundances of \hdab\ are 
summarized in Table~\ref{tab:params0607}.  The adopted line list, 
EWs, and the $\log{(N)}$ line-by-line abundances of each element for  
\hdab\ and the Sun are given in Table~\ref{tab:linelist0607}.

\section{Results\label{s:results}}
The stellar parameters (Table~\ref{tab:params0607}) determined for \hdab\ 
are consistent with both stars having a $\sim$G5V spectral type. 
The two stars are essentially stellar twins, and the differences in
the stellar parameters are on the order of the 1$-\sigma$ uncertainties:
$\Delta$\teff\ $=52 \pm 62$ K, $\Delta \log g=0.04 \pm 0.10$ dex, 
and $\Delta \xi=0.10\pm 0.10$ km s$^{-1}$}.  
The measured abundances relative to solar of the 15 individual elements
that were analyzed are summarized in Table~\ref{tab:params0607}. 

In addition, the stellar parameters of \hdab\ are similar enough that we
can perform a direct line-by-line differential abundance analysis
without using the Sun as a reference. 
\kgsinsplain{Here, we may compare the stars' abundances differences relative to the statistical 
errors alone, since the systematic uncertainties in the absolute abundance scale arising 
from the stellar parameters are eliminated by virtue of the stellar parameters being nearly identical.}
\kgsdel{On an element-by-element basis, the means of the line-by-line
differential abundances ([X/H]$_{06}-$[X/H]$_{07}$) for each element
are consistent with zero at the 3-$\sigma$ level.} 
In Figure~\ref{fig:diff0607}, for each element,  
the difference in the means of the line-by-line abundances
relative to solar are plotted.  
Averaging over all the elements measured,
the mean abundance difference is $-0.018\pm0.004$ dex \kgsdel{(\hdb\ $-$ \hda)}
\kgsinsplain{(\hda\ $-$ \hdb), where the error is the uncertainty of the mean}.
\kgsinsplain{In other words, the giant-planet hosting star (\hda) is slightly but significantly less enhanced in metals than the non-planet hosting star (\hdb).}

\kgsinsplain{As discussed in detail in Section \ref{secn:sim_acc},}
when we compare this \kgsdel{lack of a statistically significant} difference in the
abundances of \kgsdel{\hdab} \kgsinsplain{\hda$-$ \hdb\ } to the differential abundances
predicted by our simple accretion model \citep{mack14}, we find that
we can \kgsdel{place a 3-$\sigma$ upper limit of} 
\kgsins{readily exclude a scenario in which}
\kgsdel{10 \ME\ on the amount
of Earth-like (i.e., rocky) material that \hda\ could have accreted with respect to \hdb}
\kgsins{\hda\ could have accreted 10 \ME\ of Earth-like material with respect to \hdb}.
\kgsins{A scenario in which \hda\ accreted 5 \ME\ relative to \hdb\ is also strongly statistically ruled out, though we note that two of the elements (Al and Si) are not
excluded by the model.}
\kgsins{However, these two elements notwithstanding, it appears that we may 
exclude at the $\sim5\sigma$ level a scenario in which as little as $\sim$2.5 \ME\ of Earth-like material
was accreted by the planet-hosting \hda\ relative to the non-planet hosting \hdb.}
\kgsins{We discuss this result further in Section \ref{secn:sim_acc}.}

\kgsdel{Furthermore,} 
For comparison to previous studies of planet-hosting wide binaries,
we also determine how the \hdab\ abundances relative to solar ([X/H]) vary
as a function of elemental condensation temperature (\tc).
We performed both unweighted and weighted linear fits 
\kgsins{(since both types of fits have appeared in the literature)} 
to the [X/H] versus \tc\ abundance relations to investigate possible correlations.
The numerical values of the slopes derived from these fits 
are summarized in Table~\ref{tab:slopes}. 
Figure~\ref{fig:tc_weight0607} shows the
result of the weighted linear fit to the
elemental abundances of \hdab\ versus \tc.  
The condensation temperatures were taken from the 50\% \tc\ values
listed in \citet{2003ApJ...591.1220L}. 
\kgsins{Note that the \tc\ values from \citet{2003ApJ...591.1220L} are strictly speaking for solar system composition, whereas \hdab\ is metal rich with respect to solar; it is not known whether this introduces any significant additional uncertainty to the \tc\ values.}
For both the unweighted and weighted linear fits, the slopes of \hdab\ agree
within 1-$\sigma$, as they should given that we find the stars to be chemically
identical within our uncertainties. Moreover, within 2-$\sigma$, both the unweighted
and weighted slopes for both stars are consistent with zero,
except for the weighted slope of \hda, which differs from zero by 
more than 4-$\sigma$ (slope$=(-0.92\pm0.21)\times10^{-5}$~dex~K$^{-1}$).
The slightly negative slope of \hda\ indicates that it may be slightly depleted
in refractory elements relative to solar.

In summary, given that the stellar parameters for \hdab\ agree within $\sim$1$\sigma$
\kgsinsplain{(including both being overall very metal rich)},
\kgsinsplain{\hdab\ are physically similar enough that they may be considered stellar twins.
The most important distinction between these twins is that \hda\ is known to host an eccentric 
giant planet whereas \hdb\ is not known to host planets.}
The [X/H]-\tc\ slopes for \hdab\ are identical at the 1$\sigma$ level.
\kgsinsplain{However,}
the line-by-line differential abundances for each element 
\kgsdel{are consistent within zero at the 3$\sigma$ level}
\kgsinsplain{reveal that the planet-hosting \hda\ is slightly but significantly less enhanced
in metals} (Table~5), 
\kgsins{and that a scenario in which \hda\ has accreted more than $\sim$2.5 \ME\ of Earth-like material relative to \hdb\ can be excluded to high statistical significance.}
\kgsdel{\hdab\ are physically and chemically similar enough to be considered stellar twins.}

\section{Discussion\label{s:disc}}
%
%
Building on our previous work~\citep{mack14}, we again interpret our results
in terms of a simple model for the accretion of Earth-like material by a
Sun-like star. As has been suggested in both~\citet{mack14} and \citet{schuler11a},
stars with close-in giant planets may be more likely to accrete significant
amounts of rocky planetary material. This accretion, if it occurs, may be
the result of the giant planet scattering rocky planets or planetesimals into the
star as it migrates inward. 
Given that \hda\ hosts a close-in
giant planet, while \hdb\ does not, and that \hdab\ can be considered stellar twins,
then the \hdab\ system is an ideal laboratory for investigating if rocky-planet accretion
triggered by giant-planet migration occurred in this particular system. Furthermore,
systems like \hdab\ will serve as benchmark members of the growing catalog of planet-hosting
wide binaries that can be used to determine the occurrence rate of rocky-planet accretion.

\subsection{Upper limits on the amount of rocky planetary material scattered into
\hdab}
\label{secn:sim_acc}

In our analysis of \hdatetwoone\ \citep{mack14}, we created a simple model for the
impact that the accretion of a rocky planet would have on the atmospheric composition of a
solar-type star. \kgsins{Our discussion in this section refers heavily to that model; we direct the reader to \citet{mack14} for details and here describe the modifications relevant to the current analysis. In particular,} 
because the bulk compositions of \hdatetwoone\
([Fe/H]$_{82}=-0.02$ and [Fe/H]$_{81}=+0.04$) are consistent
with solar, we were able to assume in \citet{mack14} that their primordial
compositions were approximately equal to the present-day solar composition. As a result, we were able to
use literature values for the composition of the Earth~\citep{McDonough01}
and the composition of the Sun~\citep{2009ARA&A..47..481A} to estimate
how the accretion of Earth-like rocky material would alter the photospheric
abundances of a Sun-like star. Under the assumptions of our model,
the [X/H]$-$\tc\ slope for \hdatetwo\ was consistent with the ingestion of $\sim$10~$M_{\oplus}$, 
while the slope for \hdateone\ was consistent with the ingestion of $\sim$20~$M_{\oplus}$.
In both cases, this is the amount of material that \hdatetwo\ or \hdateone\ may have
accreted with respect to the Sun.  With respect to each other, their [X/H]$-$\tc\
slopes are consistent within 1-$\sigma$.

In the case of the present analysis, since the bulk compositions of \hdab~([Fe/H]$_{06}=+0.35$ and [Fe/H]$_{07}=+0.35$) are not consistent with solar, \kgsinsplain{and are in fact quite metal-rich relative to the Sun,} we need to modify our model accordingly. Instead of assuming the primordial composition of \hdab\ was consistent with solar, we assume that the primordial composition of \hdab\ was consistent with the present-day composition of \hdb. This assumption is motivated by our previous work~\citep{mack14} and \citep{schuler11a} that indicated that host stars with close-in ($\lesssim$1~AU) giant planets may exhibit [X/H]$-$\tc\ slopes that are consistent with the accretion of rocky planetary material. Therefore, since \hda\ hosts the close-in giant planet, we presume that the present-day abundances of \hdb\ are more likely to reflect the primordial compositions of both stars.

In modifying our model so that the primordial composition is consistent with \hdb, we begin with the same approach described in~\citet{mack14}. \kgsins{Namely, we determine the abundance of each element in a given amount of rocky material (in $M_{\earth}$) and in the convection zone of a star with a given mass (in $M_{\odot}$). The rocky material abundance is then added to the convection zone abundance to give a measure of how much the photospheric abundance of a given element would increase as a result of accreting a rocky planet. By varying the amount of rocky material in the model, we can simulate the accretion of rocky planets of varying masses.} {
Even though \hdab\ are quite metal-rich, given their stellar parameters we can use the relations described in~\citet{2010A&ARv..18...67T} to estimate their masses, which are consistent with solar. \citet{2001ApJ...556L..59P} showed that for stars with \teff$=5500-5600$~K, the mass in the convection zone ranges from $\sim0.03-0.04$ $M_{\odot}$ \kgsins{(here we use the most metal-rich models considered by those authors, [Fe/H]$=+0.2$; however, as those authors demonstrate, the effect on the convection zone depth of changing the metallicity by $\sim$0.1 dex is negligible)}. Therefore, in our model, it is a reasonable approximation to set the mass of \hdb\ to the mass of the Sun, and the mass of the convection zone in \hdb\ to the mass of the solar convection zone.

The results of simulating the accretion of 5 $M_{\oplus}$ and 20 $M_{\oplus}$ 
by \hdb\ are shown in Figure~\ref{fig:abd_acc07}. 
Like the planetary accretion model for solar-type stars in~\citet{mack14}, the accretion model 
for metal-rich stars like \hdb\
also predicts that the accretion of rocky planetary material with Earth-like composition
tends to create a more positive correlation between [X/H] and \tc. Increasing
the amount of accreted material causes the slopes to become more positive (or
equivalently, less negative). 
Furthermore, when comparing the results to the modified model, we can 
place upper limits on the amount of each element 
that could have been accreted by \hda\ with respect to \hdb. 


Unlike our analysis of \hdatetwoone\ \citep{mack14}, where we used the abundances relative to solar
in order to determine how much rocky material the two stars might have accreted
relative to the Sun, with \hdab, the two stars have very similar \teff\ so we can directly
compare their abundances through a differential analysis. This differential analysis
effectively eliminates the systematic uncertainties in our abundance measurements.
For this part of the analysis, we employ the
means of the line-by-line differential abundances shown in Table~4, as opposed
to the difference of the means that are given in Table~2. Since it is the
means of the line-by-line differential abundances that are important here, then
we use the uncertainty in the mean to define our precision. For most of the
elements, the uncertainty in the mean is on the order of $\sim$0.01~dex, and
all of them possess a precision $\lesssim$0.04~dex.

As a result, within the assumptions of our theoretical model, we can place tight
constraints on the amount of material that \hdab\ could have accreted with respect
to each other on an element-by-element basis. 
\kgsdel{In fact, we find that \hda\ could have
accreted at most 10 \ME\ of rocky, Earth-like material with respect to \hdb. If \hda\
had accreted more than 10 \ME\ with respect to \hdb, then we would be able to detect
this enrichment of H-depleted material at the 3$\sigma$ level for most of the
the elements that we analyzed.}
\kgsins{As shown in Figure~2 and discussed in Section~3, we can very securely rule out a
scenario in which \hda\ accreted as much as 10 \ME\ of Earth-like material relative to \hdb,
and statistically we can place an upper limit of $\sim$2.5 \ME\ (formally 10$\sigma$).}

\kgsdel{Importantly,}
This upper limit of \kgsdel{10} \kgsinsplain{2.5} \ME\ is \kgsdel{more than twice} 
\kgsinsplain{roughly half} the amount of rocky material contained in the Solar System terrestrial planets and primordial asteroid belt \citep[e.g.][]{2010ApJ...724...92C}, 
\kgsdel{This suggests that the actual amount of rocky material accreted was most likely much less than this---} and in fact the [X/H] versus \tc\ data are consistent with no rocky accretion at all \kgsinsplain{so the actual amount accreted could have been less than even this}. This implies that any primordial close-in rocky material must have been prevented from reaching the stars, as we now discuss.

\subsection{A scenario explaining the lack of a planet accretion signature in \hdab}


An important question in the study of exoplanetary systems is the frequency with which
planet-hosting stars accrete rocky planetary material 
during the formation and migration processes of giant planets.
Previous studies of planet-hosting wide binaries with FGK components where at least one component
has a close-in giant planet (by ``close-in'' we mean $\lesssim$1~AU), as well as FGK single stars
with close-in giants, have suggested that close-in giant planets may push significant amounts
of inner rocky planetary material into their host stars \citep{schuler11a,schuler11b,mack14,teske15,biazzo15}. 
If this rocky-planet accretion
process is a common occurrence, and if the rocky material is distributed evenly over the 
convection zone and remains there for a significant fraction of the star's main-sequence
lifetime, then for solar-type planet-hosting wide binaries where one component
is known to host a close-in giant planet and the other is not, the star with the
close-in giant planet, on average, may be more likely to have refractory element abundances that are
enhanced with respect to its companion. 

In the case of \hdab,
{given the fact that \hda\ has a very eccentric giant planet that approaches within 
$\lesssim$0.03 AU, it may be  surprising that its slope for [X/H] versus \tc\ does not
suggest the accretion of rocky planetary material relative to its companion or the Sun, unlike the stars in the wide binary
\hdatetwoone, which both have close-in eccentric giant planets.}
Indeed, as we have seen, there is an \kgsdel{stringent} upper limit of \kgsdel{10} \kgsinsplain{2.5} \ME\ 
for the amount of rocky material that \hda\ could have accreted 
relative to \hdb.

However, it is important to keep in mind that \kgsdel{10} \kgsinsplain{2.5}~\ME\ or more of inner rocky planets may be rare or at least uncommon. The recent discovery of Kepler-444 \citep{campante15} provides an example of a planetary system with five inner rocky planets. Had a migrating giant planet scattered them into the host star, the ingested rocky material \kgsdel{would have been} 
\kgsinsplain{may} well \kgsinsplain{have been} below our current detection threshold of \kgsdel{10} \kgsinsplain{2.5}~\ME\ for the \kgsdel{\hdab\ system} \kgsinsplain{\hda}. In the case of \hda, if it had ingested four planets identical in mass to the inner rocky planets in the Solar System (and with our model assumption that the planets have Earth-like composition), we would have been unable to detect the increase in the rocky elemental abundances. Therefore, the question of how often a rocky-planet accretion event occurs for solar-type stars can only truly be answered as a function of accreted mass, and our data for the \hdab\ system cannot provide a clear picture for ingested masses below \kgsdel{10} \kgsinsplain{2.5}~\ME.

{Finally, a recent study by Mustill et al.\ (2015) carried out numerical simulations of giant planets on eccentric orbits with pericenters of $\lesssim{3}$ AU. That study found that not only did these
giant planets frequently interact strongly with inner rocky planets, but the giant planet also could in some cases accrete the rocky planets onto themselves. 
This would lead to giant planets that have cores enriched in heavy elements, such as those studied by Guillot et al.\ (2006). In this case, it would be interesting to investigate if \hda b
shows evidence of having an enriched core, which could potentially explain why \hda\ failed to accrete more than \kgsdel{10} \kgsinsplain{2.5}~\ME\ of rocky material relative to \hdb.}


\subsection{Comparison to Prior Work}


{Our results are in good agreement with the recent findings of \hdab\ by \citet{saffe15}. They also found that \hdab\ were not depleted in refractory elements. As they note, the lack of a trend with \tc\ does appear to imply that the existence of a close-in giant planet does not always lead to the ingestion of H-depleted, refractory-rich planetary material by the host star \citep[a conclusion that also agrees with the result from][]{lui14}.}

{Our own analysis extends the \citet{saffe15} findings by adding the quantitative model of \citet{mack14}, which allows us to stringently constrain the amount of rocky material that could have been ingested by \kgsdel{\hdab\ relative to each other} 
\kgsinsplain{the planet-hosting \hda\ relative to the non-planet-hosting \hdb}: $\lesssim$\kgsdel{10} \kgsinsplain{2.5}~\ME. 
Applying our model to the individual abundance measurements reported by \citet{saffe15}, we find, given the average differential abundance of $+0.010\pm0.019$ dex reported by them, a similar upper limit of 
$\sim$5 \ME\ (3$\sigma$ confidence)
\kgsins{although with less precision than the upper limit derived from our work reported here}. 
In other words, 
based on the level of differential abundances that can be reliably measured by present-day techniques such as used in \citet{saffe15} and in the present study, 
it remains possible that
\hda\ could have ingested \kgsdel{$\sim5$--10} \kgsinsplain{2.5--5} \ME\ with respect to \hdb. 
\kgsdel{Again, we point out that even 5 \ME\ is more rocky material than contained in the Solar System planets and primordial asteroid belt.}
}

{At the same time, it remains true that chemical abundance studies of PHWBs with close-in giant planets have found varying results. Some appear to see a clear enhancement of at least a few rocky elements when performing a differential abundance analysis between the two stars in a given PHWB, such as the enhancement in refractory elements of X0-2N with respect to X0-2S reported by \citet{teske15}, \citet{biazzo15} and \citet{2015ApJ...808...13R}, as well as the enrichment of 16 Cyg B with respect to 16 Cyg A reported by \citet{ramirez11} and \citet{tucci14}. On the other hand, others have found essentially no difference between the chemical abundances of the stars in the PHWB: \citet{saffe15} and this present work, the study of HAT-P-1 by \citet{lui14}, and a separate analysis of the 16 Cyg system performed \citet{schuler11b}.} \kgsins{Also, while not a direct measure of the elemental abundances of the stellar photospheres, asteroseismic modeling of 16 Cyg A and B suggests that the two stars have the same chemical compositions \citep{metcalfe12,metcalfe15}.}

{These mixed results underscore the need for an analysis of a larger sample of PHWBs and 
for analysis by multiple investigators using different instruments and analysis techniques. 
\kgsinsplain{Moreover, increasing the sample size of detailed abundances of PHWBs will facilitate the more precise delineation of possible chemical signatures resulting from the accretion of rocky planet material and planet formation itself, and the disentanglement of the effects of the two processes.} 
\kgsins{Whereas the accretion of H-depleted rocky planet material would be expected to enhance the abundances of refractory elements of a star, it has been suggested that planet formation, rocky planet formation in particular, may indeed deplete refractory abundances of a star \citep[e.g.][]{2009ApJ...704L..66M,2009A&A...508L..17R}, although not all observational results support this suggestion \citep[e.g.][]{2014A&A...564L..15A,2015ApJ...815....5S}.} 
\kgsinsplain{The available evidence across the various PHWBs that have been analyzed strongly indicates real astrophysical variation in the abundance patterns of stars with planets, and the reasons for the variation, whether due to accretion, planet formation, Galactic chemical evolution, or some other unknown effect, remain unclear.}



\section{Summary and Conclusion\label{s:conc}}
We have used Keck/HIRES observations to perform a detailed spectroscopic analysis 
of the binary \hdab, in which one star (\hda) is known to host a very eccentric
($e \approx 0.93$) giant planet with closest approach of $\sim$0.03 AU,
while its sibling (\hdb) is not known to host planets.
We find that the two stars are otherwise ``twin" G5 dwarfs, with nearly identical physical
properties, and presumably with the same age and initial chemical makeup, making this an 
unusually good astrophysical laboratory for exploring 
differences in planet hosting characteristics among otherwise similar stars.

Following our methodology from analysis of another PHWB system
\citep[\hdatetwoone;][]{mack14},
we have performed a detailed chemical abundance analysis of the two stars in \hdab.
\kgsins{Similar to the recent results of \citet{saffe15}, we find linear least-squares fits
to the refractory elemental abundances ([X/H]) versus elemental
condensation temperature (\tc) yield slopes that are statistically
indistinguishable for the two stars. Evidently, the very different planetary system
architectures of the two stars did not result in a difference greater than
$\sim$0.03--0.04 dex in the photospheric elemental
abundances of the stars.} This finding is in stark contrast to the clearly different
[X/H] vs.\ \tc\ patterns that we found for the two stars in \hdatetwoone\
\citep{mack14}.
Indeed, in that analysis we developed a simple model for the pollution of the
stellar photospheres by ingestion of small rocky planets, likely previously scattered 
into the stars by the currently observed giant planets, which could successfully 
explain the observed chemical abundance patterns and differences between the two stars.

\kgsinsplain{However, when we analyze the abundance differences on a relative,} 
element-by-element basis, \kgsinsplain{thus eliminating potential systematics arising from 
uncertainties in the stellar parameters,} the means of the line-by-line differential abundances ([X/H]$_{06}-$[X/H]$_{07}$) for each element
\kgsinsplain{show that the heavy element abundances of \hda\ are $-0.018\pm0.004$ dex relative to those of \hdb}.
\kgsdel{are consistent with zero at the 3$\sigma$ level.}
In the present case of \hdab, our rocky planet accretion model suggests an upper limit of 10~\ME\ for the differential amount of rocky planetary material that could have scattered onto \hda\ over \hdb\ \kgsdel{or vice-versa}.
In fact the observations are consistent with zero rocky planetary accretion. 
Indeed, even \kgsdel{10} \kgsinsplain{2.5}~\ME\ would represent the accretion of \kgsdel{more than twice} \kgsinsplain{about half} the amount of rocky material sequestered in the terrestrial planet of the Solar System, plus the primordial asteroid belt. Thus we consider it most likely that 
\kgsdel{\hdab} \kgsinsplain{\hda} accreted little if any rocky planetary material \kgsinsplain{relative to \hdb}.
This may seem surprising particularly in the case of \hda, given the fact that it hosts a very eccentric giant planet that approaches within $\lesssim$0.03 AU, and thus might be expected to very easily sweep inner rocky planets into the star. 


However, as we have discussed, if the giant planet in \hda\ previously migrated 
inward very quickly, it may have been an inefficient
shepherd of material interior to its orbit. 
We suggest that this idea is plausible
given that larger orbital eccentricities are often the result of a more violent migration
history, such as planet-planet scattering.
\kgsinsplain{In this scenario, we expect that any rocky planets that might have been originally interior to the 
\hda\ giant planet would have been preferentially ejected from the 
system, rather than directed into the host star.}

\acknowledgements
C.E.M. and K.G.S acknowledge support by {NSF AAG AST-1009810 and NSF PAARE AST-0849736}. S.C.S. acknowledges support provided by grant NNX12AD19G from the National Aeronautical and Space Administration as part of the Kepler Participating Scientist Program.
The data presented herein were obtained at the W.M. Keck Observatory, which is operated as a scientific partnership among the California Institute of Technology, the University of California and the National Aeronautics and Space Administration. The Observatory was made possible by the generous financial support of the W.M. Keck Foundation. 
The authors wish to recognize and acknowledge the very significant cultural role and reverence that the summit of Mauna Kea has always had within the indigenous Hawaiian community.  We are most fortunate to have the opportunity to conduct observations from this mountain.
 
{\it Facilities:} \facility{Keck I: HIRES}



\begin{deluxetable}{lcccccccc}
\tablecolumns{9}
\tablewidth{0pt}
\tablecaption{Stellar Parameters \& Abundances\tablenotemark{a}\label{tab:params0607}}
\tablehead{
	\colhead{}&
	\colhead{}&
	\colhead{\hda}&
	\colhead{}&
	\colhead{\hdb}&
	\colhead{}&
	\colhead{\hda$-$\hdb}&
	\colhead{}&
	\colhead{\tc\ (K)}
	}
\startdata
$T_{\mathrm{eff}}$ (K) && $5613 \pm 44$          && $5561 \pm 43$   && $52 \pm 62$ && \nodata \\
$\log g$ (cgs)         && $4.43 \pm 0.08$        && $4.47 \pm 0.06$ && $0.04 \pm 0.10$ && \nodata \\
$\xi$ (km s$^{-1}$)    && $1.36 \pm 0.07$        && $1.26 \pm 0.07$ && $0.10 \pm 0.10$ && \nodata \\
$[$Na/H$]$  \dotfill   && $+0.45 \pm 0.03 \, \pm 0.05$ && $+0.46 \pm0.04 \, \pm 0.05$ && $-0.01 \pm0.01 \, \pm 0.02$ && 958 \\
$[$Mg/H$]$\tablenotemark{d} \dotfill   && $+0.40 \pm 0.00 \, \pm 0.03$ && $+0.47 \pm0.00 \, \pm 0.03$ && \nodata && 1336 \\
$[$Al/H$]$  \dotfill   && $+0.39 \pm 0.01 \, \pm 0.03$ && $+0.37 \pm0.02 \, \pm 0.03$ && $+0.02 \pm0.01 \, \pm 0.01$ && 1653 \\
$[$Si/H$]$  \dotfill   && $+0.34 \pm 0.01 \, \pm 0.01$ && $+0.34 \pm0.01 \, \pm 0.01$ && $+0.01 \pm0.01 \, \pm 0.02$ && 1310 \\
$[$Ca/H$]$  \dotfill   && $+0.28 \pm 0.01 \, \pm 0.05$ && $+0.31 \pm0.01 \, \pm 0.04$ && $-0.03 \pm0.00 \, \pm 0.01$ && 1517 \\
$[$Sc/H$]$  \dotfill   && $+0.36 \pm 0.02 \, \pm 0.04$ && $+0.39 \pm0.04 \, \pm 0.05$ && $-0.03 \pm0.02 \, \pm 0.03$ && 1659 \\
$[$Ti/H$]$  \dotfill   && $+0.36 \pm 0.02 \, \pm 0.05$ && $+0.42 \pm0.02 \, \pm 0.05$ && $-0.05 \pm0.02 \, \pm 0.05$ && 1582 \\
$[$V/H$]$   \dotfill   && $+0.41 \pm 0.01 \, \pm 0.05$ && $+0.45 \pm0.02 \, \pm 0.06$ && $-0.03 \pm0.01 \, \pm 0.03$ && 1429 \\
$[$Cr/H$]$  \dotfill   && $+0.34 \pm 0.04 \, \pm 0.06$ && $+0.39 \pm0.04 \, \pm 0.05$ && $-0.02 \pm0.02 \, \pm 0.04$ && 1296 \\
$[$Mn/H$]$\tablenotemark{d}  \dotfill   && $+0.42 \pm 0.00 \, \pm 0.06$ && $+0.39 \pm0.00 \, \pm 0.06$ && \nodata && 1158 \\
$[$Fe/H$]$  \dotfill   && $+0.35 \pm 0.01 \, \pm 0.02$ && $+0.35 \pm0.01 \, \pm 0.02$ && $+0.00 \pm0.01 \, \pm 0.04$ && 1334 \\
$[$Co/H$]$  \dotfill   && $+0.44 \pm 0.01 \, \pm 0.04$ && $+0.46 \pm0.01 \, \pm 0.04$ && $-0.03 \pm0.01 \, \pm 0.02$ && 1352 \\
$[$Ni/H$]$  \dotfill   && $+0.40 \pm 0.01 \, \pm 0.03$ && $+0.43 \pm0.01 \, \pm 0.03$ && $-0.03 \pm0.01 \, \pm 0.03$ && 1353 \\
\enddata
\tablenotetext{a}{Adopted solar parameters: \teff\ $=5777$ K, $\log g=4.44$, 
and $\xi=1.38$ km s$^{-1}$.}
\tablenotetext{b}{$\sigma_{\mu}$ -- the uncertainty in the mean}
\tablenotetext{c}{$\sigma_{Total}$-- quadratic sum of $\sigma_{\mu}$ and uncertainties due to uncertainties in \teff, $\log g$, and $\xi$.}
\tablenotetext{d}{In both stars, the abundance measurements for Mg and Mn were determined from a single spectral line.  That is why the uncertainty in the mean is 0.00 in both stars.}
\end{deluxetable}

\begin{deluxetable}{ccc}
\tablecaption{Unweighted and Weighted Linear Fits to [X/H] vs $T_{\rm eff}$ \label{tab:slopes}}
\tablecolumns{3}
\tablewidth{0pt}
\tablehead{
  \colhead{} &
  \colhead{Unweighted Slope} &
  \colhead{Weighted Slope} \\
  \colhead{} &
  \colhead{(dex K$^{-1}$)} &
  \colhead{(dex K$^{-1}$)}
}
\startdata
\hda\ & $(-11.56\pm6.40)\times10^{-5}$ & $(-0.92\pm0.21)\times10^{-5}$ \\
\hdb\ & $(-8.31\pm7.45)\times10^{-5}$ & $(-0.54\pm0.30)\times10^{-5}$ \\
\enddata
\end{deluxetable}

\newpage
\setlength{\tabcolsep}{1mm}
\begin{deluxetable}{lcccccccrcccrcccrcc}
\tablecolumns{19}
\tablewidth{0pt}
\tabletypesize{\scriptsize}
\tablecaption{\hdab : Lines Measured, Equivalent Widths, and Abundances\label{tab:linelist0607}}
\tablehead{
     \colhead{}&
     \colhead{}&
     \colhead{$\lambda$}&
     \colhead{}&
     \colhead{$\chi$}&
     \colhead{}&
     \colhead{}&
     \colhead{}&
     \colhead{}&
     \colhead{}&
     \colhead{}&
     \colhead{}&
     \multicolumn{3}{c}{\hda}&
     \colhead{}&
     \multicolumn{3}{c}{\hdb}\\ 
     \cline{13-15} \cline{17-19}\\
     \colhead{Element}&
     \colhead{}&
     \colhead{(\AA)}&
     \colhead{}&
     \colhead{(eV)}&
     \colhead{}&
     \colhead{$\log \mathrm{gf}$}&
     \colhead{}&
     \colhead{EW$_{\odot}$}&
     \colhead{$\log N_{\odot}$}&
     \colhead{$\log N_{\odot,\mathrm{synth}}$\tablenotemark{a}}&
     \colhead{}&
     \colhead{EW}&
     \colhead{$\log N$}&
     \colhead{$\log N_{\mathrm{synth}}$}&
     \colhead{}&
     \colhead{EW}&
     \colhead{$\log N$}&
     \colhead{$\log N_{\mathrm{synth}}$}\\
      }
\startdata
\ion{Na}{1} && 5682.63 && 2.10 && -0.700 && 100.9 & 6.30 & \nodata && 142.9 & 6.70 & \nodata && 146.0 & 6.70 & \nodata \\ 
\ion{Na}{1} && 6154.23 && 2.10 && -1.560 &&  38.4 & 6.31 & \nodata &&  76.5 & 6.77 & \nodata &&  80.4 & 6.80 & \nodata \\ 
\ion{Na}{1} && 6160.75 && 2.10 && -1.260 &&  56.1 & 6.28 & \nodata && 100.6 & 6.78 & \nodata && 102.1 & 6.77 & \nodata \\ 
\ion{Mg}{1} && 4730.03 && 4.35 && -2.523 &&  67.3 & 7.83 & \nodata && 103.0 & 8.23 & \nodata && 110.2 & 8.30 & \nodata \\ 
\ion{Al}{1} && 6696.02 && 3.14 && -1.347 &&  38.3 & 6.27 & \nodata &&  70.4 & 6.66 & \nodata &&  70.8 & 6.65 & \nodata \\ 
\ion{Al}{1} && 6698.67 && 3.14 && -1.647 &&  22.9 & 6.27 & \nodata &&  47.9 & 6.65 & \nodata &&  47.9 & 6.62 & \nodata \\ 
\ion{Si}{1} && 5701.10 && 4.93 && -1.581 &&  37.2 & 7.08 & \nodata &&  59.8 & 7.46 & \nodata &&  57.6 & 7.44 & \nodata \\ 
\ion{Si}{1} && 5772.15 && 5.08 && -1.358 &&  51.8 & 7.23 & \nodata &&  72.6 & 7.56 & \nodata &&  71.8 & 7.56 & \nodata \\ 
\ion{Si}{1} && 6125.02 && 5.61 && -1.464 &&  30.3 & 7.44 & \nodata &&  52.2 & 7.83 & \nodata &&  50.5 & 7.81 & \nodata \\ 
\ion{Si}{1} && 6142.48 && 5.62 && -1.295 &&  32.4 & 7.31 & \nodata &&  52.8 & 7.67 & \nodata &&  53.2 & 7.69 & \nodata \\ 
\ion{Si}{1} && 6145.02 && 5.62 && -1.310 &&  37.2 & 7.41 & \nodata &&  58.4 & 7.76 & \nodata &&  56.0 & 7.74 & \nodata \\ 
\ion{Si}{1} && 6243.81 && 5.62 && -1.242 &&  47.0 & 7.50 & \nodata &&  67.9 & 7.83 & \nodata &&  68.2 & 7.84 & \nodata \\ 
\enddata
\tablecomments{This table is shown in its entirety on the last three pages of this manuscript (Table~\ref{tab:linelist_full0607}).  
A portion is shown here for guidance regarding 
its form and content.}
\tablenotetext{a}{Indicates the $\log{N}$ abundance determined from the synthetic fit to a given line. Each synthetic
                  fit was performed with the {\sc moog} {\it synth} driver. Synthetic fits were only 
                  performed for the subset of V, Mn, and Co lines that were tested for hfs.}
\tablenotetext{c}{Indicates that the spectral line was tested for hfs.}
\end{deluxetable}

\newpage
\begin{deluxetable}{lccccccc}
\tablecolumns{8}
\tablewidth{0pt}
\tablecaption{Comparison of the observed and predicted differential abundances\protect\\ for various amounts of ingested Earth-masses\label{tab:obs_vs_preds}}
\tablehead{
     \colhead{}&
     \colhead{}&
     \colhead{\underline{Observed $\Delta$[X/H]}}&
     \colhead{}&
     \colhead{}&
     \colhead{\underline{Predicted $\Delta$[X/H]}}&
     \colhead{}&
     \colhead{}\\
     \colhead{}&
     \colhead{}&
     \colhead{}&
     \colhead{}&
     \colhead{}&
     \colhead{(if \hda\ ingested $x$\ME\ w.r.t \hdb)}&
     \colhead{}& 
     \colhead{}\\
     \colhead{$\Delta$[X/H]}&
     \colhead{}&
     \colhead{\hda\ $-$ \hdb}&
     \colhead{}&
     \colhead{2.5$M_{\oplus}$}&
     \colhead{5$M_{\oplus}$}&
     \colhead{10$M_{\oplus}$}
}
%
%
\startdata
$\Delta$[Na/H] && $-0.01\pm0.01$\tablenotemark{a} && +0.002 & +0.004 & +0.009\\
$\Delta$[Mg/H] && \nodata && +0.007 & +0.015 & +0.029\\
$\Delta$[Al/H] && $+0.02\pm0.01$ && +0.012 & +0.024 & +0.047\\
$\Delta$[Si/H] && $+0.01\pm0.01$ && +0.011 & +0.022 & +0.043\\
$\Delta$[Ca/H] && $-0.03\pm0.01$ && +0.013 & +0.026 & +0.050\\
$\Delta$[Sc/H] && $-0.03\pm0.02$ && +0.010 & +0.019 & +0.037\\
$\Delta$[Ti/H] && $-0.05\pm0.02$ && +0.010 & +0.020 & +0.038\\
$\Delta$[V/H]  && $-0.03\pm0.01$ && +0.011 & +0.023 & +0.045\\
$\Delta$[Cr/H] && $-0.02\pm0.02$ && +0.011 & +0.023 & +0.044\\
$\Delta$[Mn/H] && \nodata && +0.006 & +0.013 & +0.025\\
$\Delta$[Fe/H] && $+0.00\pm0.01$ && +0.006 & +0.011 & +0.022\\
$\Delta$[Co/H] && $-0.03\pm0.01$ && +0.007 & +0.014 & +0.028\\
$\Delta$[Ni/H] && $-0.03\pm0.01$ && +0.010 & +0.019 & +0.037\\ 
\enddata
\tablenotetext{a}{The error in the mean of the line-by-line differential abundances.}
\tablecomments{The predicted $\Delta$[X/H] assumes that the observed chemical composition of \hdb\ is equivalent
to the primordial composition of both \hdab. The predicted values show how the abundances of \hdb\ would
change after accreting varying amounts ($X_{1}\,$\ME) of material with Earth-like composition, i.e., the differential abundance that
we would expect if \hda\ had accreted $X_{1}\,$\ME while \hdb\ accreted nothing.}
\end{deluxetable}


%
\newpage
\begin{figure}[H]
\centering
\includegraphics[scale=0.9]{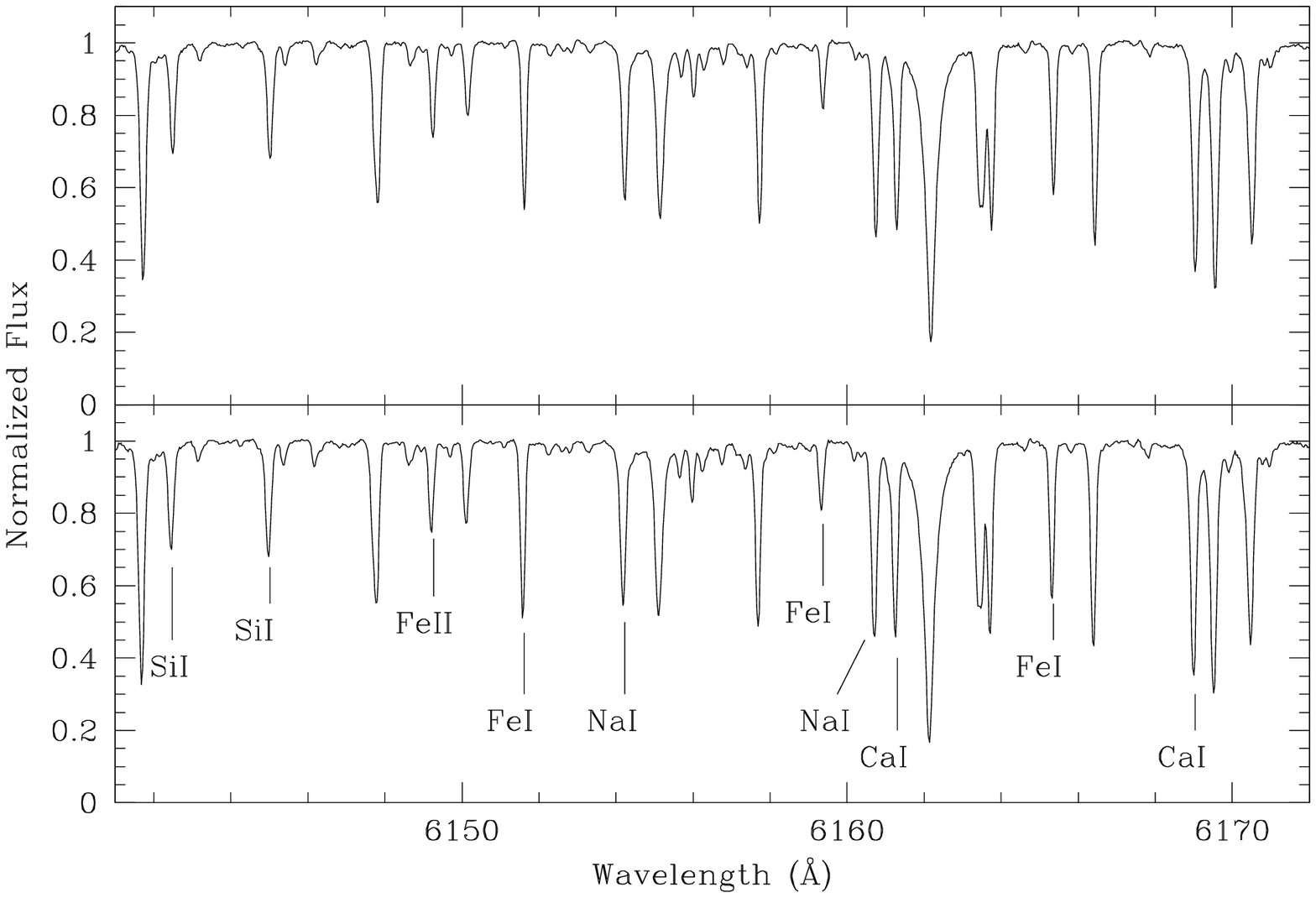}
\caption{Sample Keck/HIRES spectra for \hdab, spanning the wavelength range from $\sim\lambda6140-\lambda6170$. The marked lines are a subset of those that were measured.}
\label{fig:spec0607}
\end{figure}
%

%
\newpage
\begin{figure}[H]
\centering
\includegraphics[scale=0.7,angle=90]{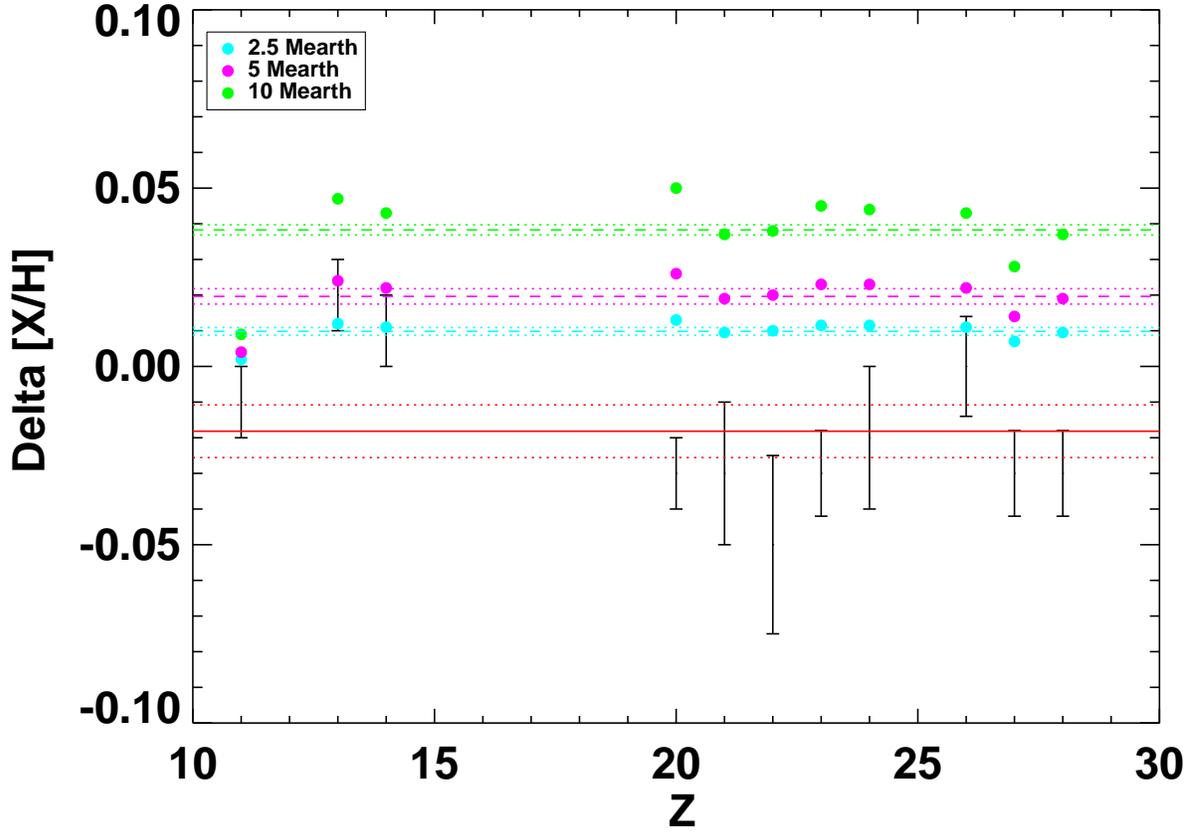}
\caption{Differential abundances for \hda$-$\hdb\ as a function of atomic number (Z).
The solid \kgsins{red} line represents the mean difference of $-0.018\pm0.004$ dex,
and the dashed line \kgsdel{is meant to guide the eye at 0.00 dex}
\kgsins{represent the model predictions of Mack et al.\ (2014) for different amounts of rocky material ingestion by the planet-hosting \hda\ relative to non-planet-hosting \hdb.}}
\label{fig:diff0607}
\end{figure}
%

%
\newpage
\begin{figure}[H]
\centering
\includegraphics[scale=0.9]{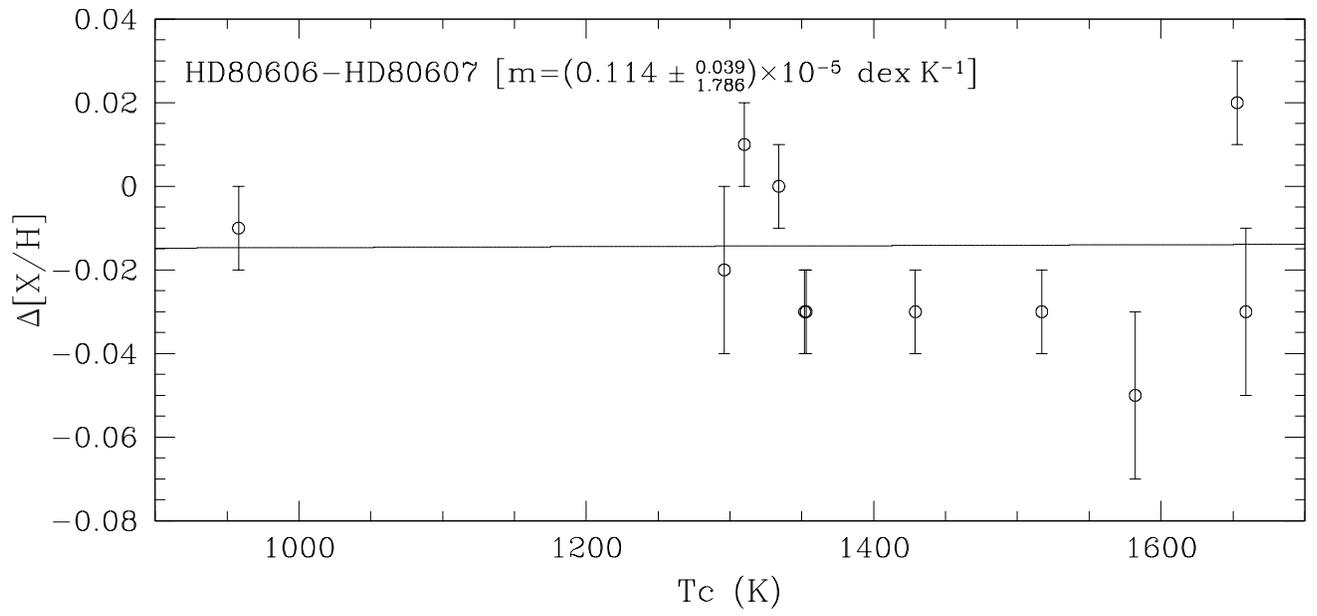}
\caption{Differential abundances for \hdab\ as a function of condensation temperature (\tc).}
\label{fig:diff0607_Tc}
\end{figure}
%


%
\newpage
\begin{figure}
\centering
\includegraphics[scale=0.85]{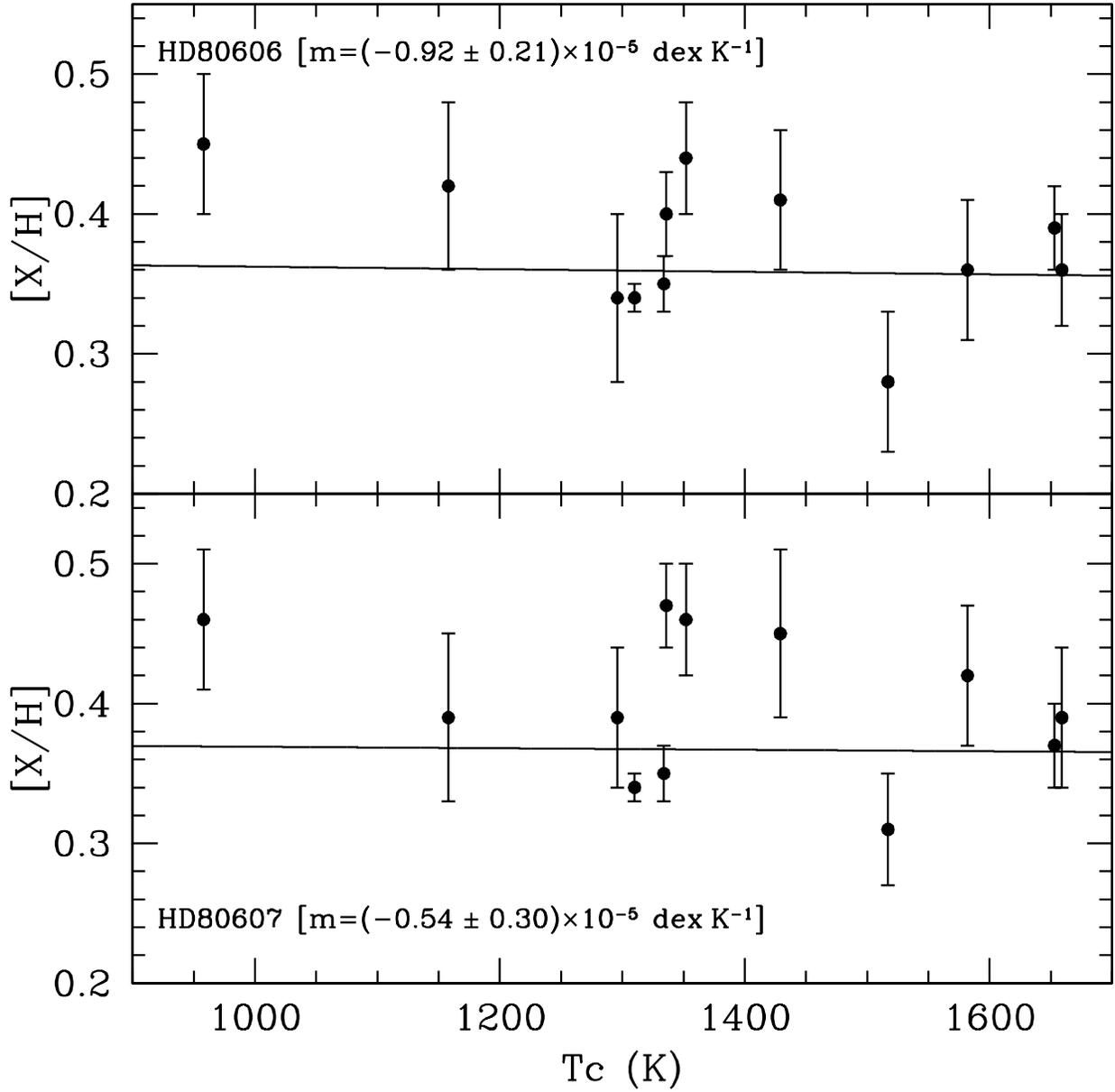}
\caption{\emph{Weighted} linear fits to abundances vs. condensation temperature (\tc) for \hdab.}
\label{fig:tc_weight0607} 
\end{figure}

%
\newpage
\begin{figure}
\centering
\includegraphics[scale=0.85]{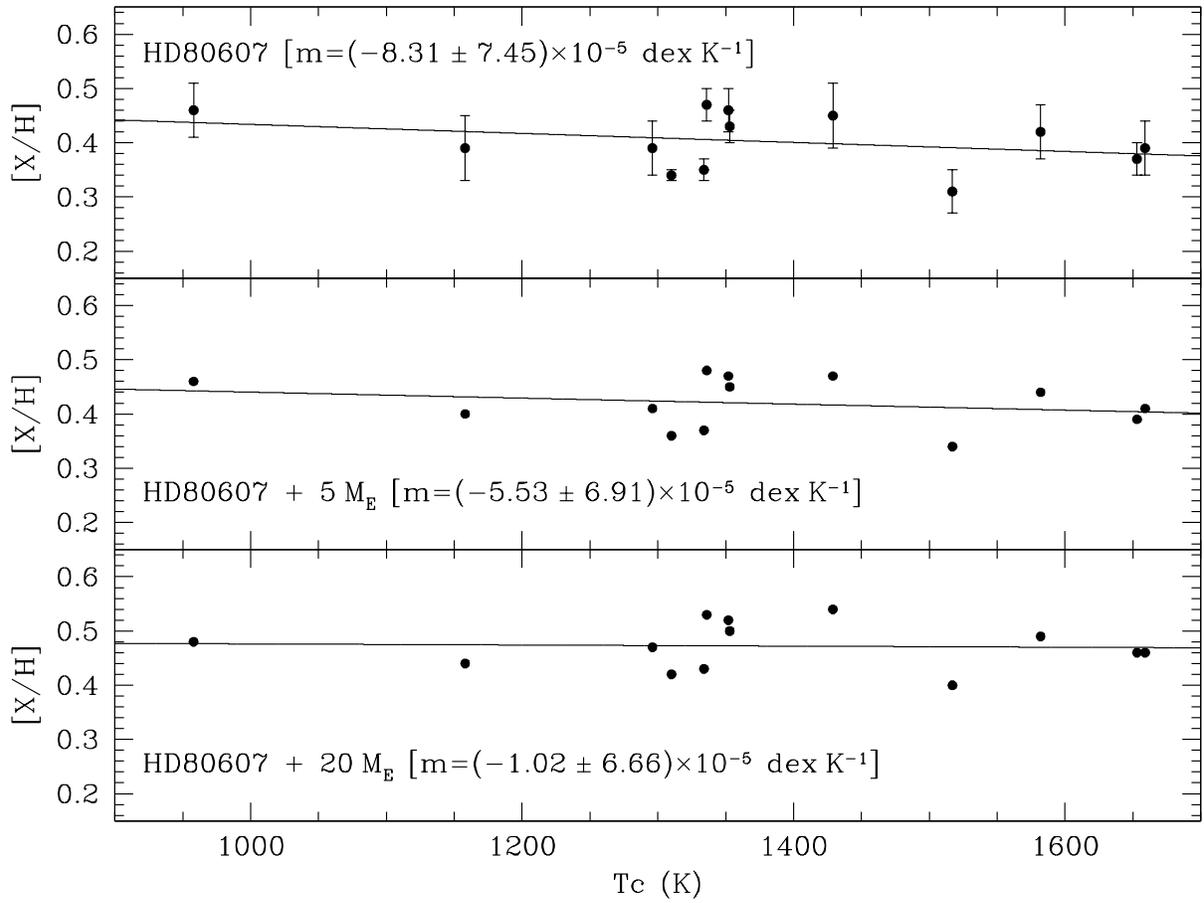}
\caption{{\bf Top:} \emph{Unweighted} linear fit to \hdb. {\bf Middle and Bottom:} \emph{Unweighted} linear fits to {simulated} abundances vs.\ condensation temperature (\tc) from our {modeled} 
accretion of 5\ME\ and 20\ME\ of material with Earth-like composition by a star like HD\,80607.}
\label{fig:abd_acc07} 
\end{figure}
%
%

%
\newpage
\setlength{\tabcolsep}{1mm}
\begin{deluxetable}{lcccccccrcccrcccrcc}
\tablecolumns{19}
\tablewidth{0pt}
\tabletypesize{\scriptsize}
\tablecaption{\hdab : Lines Measured, Equivalent Widths, and Abundances\label{tab:linelist_full0607}}
\tablehead{
     \colhead{}&
     \colhead{}&
     \colhead{$\lambda$}&
     \colhead{}&
     \colhead{$\chi$}&
     \colhead{}&
     \colhead{}&
     \colhead{}&
     \colhead{}&
     \colhead{}&
     \colhead{}&
     \colhead{}&
     \multicolumn{3}{c}{\hda}&
     \colhead{}&
     \multicolumn{3}{c}{\hdb}\\ 
     \cline{13-15} \cline{17-19}\\
     \colhead{Element}&
     \colhead{}&
     \colhead{(\AA)}&
     \colhead{}&
     \colhead{(eV)}&
     \colhead{}&
     \colhead{$\log \mathrm{gf}$}&
     \colhead{}&
     \colhead{EW$_{\odot}$}&
     \colhead{$\log N_{\odot}$}&
     \colhead{$\log N_{\odot,\mathrm{synth}}$\tablenotemark{a}}&
     \colhead{}&
     \colhead{EW}&
     \colhead{$\log N$}&
     \colhead{$\log N_{\mathrm{synth}}$}&
     \colhead{}&
     \colhead{EW}&
     \colhead{$\log N$}&
     \colhead{$\log N_{\mathrm{synth}}$}\\
      }
\startdata
\ion{Na}{1} && 5682.63 && 2.10 && -0.700 && 100.9 & 6.30 & \nodata && 142.9 & 6.70 & \nodata && 146.0 & 6.70 & \nodata \\ 
\ion{Na}{1} && 6154.23 && 2.10 && -1.560 &&  38.4 & 6.31 & \nodata &&  76.5 & 6.77 & \nodata &&  80.4 & 6.80 & \nodata \\ 
\ion{Na}{1} && 6160.75 && 2.10 && -1.260 &&  56.1 & 6.28 & \nodata && 100.6 & 6.78 & \nodata && 102.1 & 6.77 & \nodata \\ 
\ion{Mg}{1} && 4730.03 && 4.35 && -2.523 &&  67.3 & 7.83 & \nodata && 103.0 & 8.23 & \nodata && 110.2 & 8.30 & \nodata \\ 
\ion{Al}{1} && 6696.02 && 3.14 && -1.347 &&  38.3 & 6.27 & \nodata &&  70.4 & 6.66 & \nodata &&  70.8 & 6.65 & \nodata \\ 
\ion{Al}{1} && 6698.67 && 3.14 && -1.647 &&  22.9 & 6.27 & \nodata &&  47.9 & 6.65 & \nodata &&  47.9 & 6.62 & \nodata \\ 
\ion{Si}{1} && 5701.10 && 4.93 && -1.581 &&  37.2 & 7.08 & \nodata &&  59.8 & 7.46 & \nodata &&  57.6 & 7.44 & \nodata \\ 
\ion{Si}{1} && 5772.15 && 5.08 && -1.358 &&  51.8 & 7.23 & \nodata &&  72.6 & 7.56 & \nodata &&  71.8 & 7.56 & \nodata \\ 
\ion{Si}{1} && 6125.02 && 5.61 && -1.464 &&  30.3 & 7.44 & \nodata &&  52.2 & 7.83 & \nodata &&  50.5 & 7.81 & \nodata \\ 
\ion{Si}{1} && 6142.48 && 5.62 && -1.295 &&  32.4 & 7.31 & \nodata &&  52.8 & 7.67 & \nodata &&  53.2 & 7.69 & \nodata \\ 
\ion{Si}{1} && 6145.02 && 5.62 && -1.310 &&  37.2 & 7.41 & \nodata &&  58.4 & 7.76 & \nodata &&  56.0 & 7.74 & \nodata \\ 
\ion{Si}{1} && 6243.81 && 5.62 && -1.242 &&  47.0 & 7.50 & \nodata &&  67.9 & 7.83 & \nodata &&  68.2 & 7.84 & \nodata \\ 
\ion{Si}{1} && 6244.47 && 5.62 && -1.093 &&  47.3 & 7.35 & \nodata &&  67.5 & 7.67 & \nodata &&  67.7 & 7.68 & \nodata \\ 
\ion{Si}{1} && 6414.98 && 5.87 && -1.035 &&  51.5 & 7.55 & \nodata &&  71.2 & 7.83 & \nodata &&  70.3 & 7.82 & \nodata \\ 
\ion{Si}{1} && 6741.63 && 5.98 && -1.428 &&  15.2 & 7.35 & \nodata &&  29.1 & 7.72 & \nodata &&  29.0 & 7.72 & \nodata \\ 
\ion{Si}{1} && 6848.58 && 5.86 && -1.524 &&  17.5 & 7.41 & \nodata &&  32.7 & 7.78 & \nodata &&  30.6 & 7.76 & \nodata \\ 
\ion{Si}{1} && 7405.77 && 5.61 && -0.313 &&  97.9 & 7.19 & \nodata && 115.8 & 7.44 & \nodata && \nodata & \nodata & \nodata \\ 
\ion{Ca}{1} && 6161.30 && 2.52 && -1.266 &&  63.5 & 6.30 & \nodata &&  90.5 & 6.62 & \nodata &&  92.7 & 6.64 & \nodata \\ 
\ion{Ca}{1} && 6166.44 && 2.52 && -1.142 &&  67.2 & 6.24 & \nodata &&  93.6 & 6.54 & \nodata &&  96.0 & 6.56 & \nodata \\ 
\ion{Ca}{1} && 6169.04 && 2.52 && -0.797 &&  89.4 & 6.23 & \nodata && 118.7 & 6.55 & \nodata && 123.3 & 6.58 & \nodata \\ 
\ion{Ca}{1} && 6169.56 && 2.53 && -0.478 && 107.7 & 6.18 & \nodata && 138.9 & 6.47 & \nodata && 145.4 & 6.51 & \nodata \\ 
\ion{Ca}{1} && 6455.60 && 2.52 && -1.340 &&  60.6 & 6.32 & \nodata &&  81.6 & 6.54 & \nodata &&  84.3 & 6.56 & \nodata \\ 
\ion{Ca}{1} && 6499.65 && 2.52 && -0.818 &&  86.5 & 6.19 & \nodata && 108.8 & 6.41 & \nodata && 112.5 & 6.44 & \nodata \\ 
\ion{Sc}{2} && 6245.64 && 1.51 && -1.030 &&  34.7 & 3.07 & \nodata &&  48.9 & 3.44 & \nodata &&  49.4 & 3.49 & \nodata \\ 
\ion{Sc}{2} && 6604.60 && 1.36 && -1.309 &&  38.5 & 3.25 & \nodata &&  51.3 & 3.59 & \nodata &&  50.1 & 3.60 & \nodata \\ 
\ion{Ti}{1} && 5022.87 && 0.83 && -0.434 &&  69.5 & 4.74 & \nodata &&  94.2 & 5.04 & \nodata && 101.2 & 5.17 & \nodata \\ 
\ion{Ti}{1} && 5024.84 && 0.82 && -0.602 &&  66.4 & 4.84 & \nodata &&  97.0 & 5.26 & \nodata &&  99.7 & 5.30 & \nodata \\ 
\ion{Ti}{1} && 5039.96 && 0.02 && -1.130 &&  69.8 & 4.63 & \nodata &&  99.2 & 5.00 & \nodata && 105.6 & 5.12 & \nodata \\ 
\ion{Ti}{1} && 5866.45 && 1.07 && -0.840 &&  48.2 & 4.92 & \nodata &&  77.3 & 5.25 & \nodata &&  82.0 & 5.31 & \nodata \\ 
\ion{Ti}{1} && 6091.17 && 2.27 && -0.423 &&  15.5 & 4.97 & \nodata &&  33.0 & 5.27 & \nodata &&  36.2 & 5.29 & \nodata \\ 
\ion{Ti}{1} && 6098.66 && 3.06 && -0.010 &&   5.0 & 4.78 & \nodata &&  15.2 & 5.21 & \nodata &&  16.9 & 5.23 & \nodata \\ 
\ion{Ti}{1} && 6258.10 && 1.44 && -0.355 &&  49.2 & 4.80 & \nodata &&  77.7 & 5.12 & \nodata &&  80.8 & 5.15 & \nodata \\ 
\ion{Ti}{2} && 5336.79 && 1.58 && -1.590 &&  68.5 & 4.78 & \nodata &&  81.0 & 5.13 & \nodata &&  78.9 & 5.14 & \nodata \\ 
\ion{Ti}{2} && 5381.02 && 1.57 && -1.920 &&  56.4 & 4.86 & \nodata &&  73.4 & 5.29 & \nodata &&  73.8 & 5.35 & \nodata \\ 
\ion{V}{1}  && 6081.44 && 1.05 && -0.579 &&  14.0 & 3.88 & \nodata &&  38.8 & 4.29 & \nodata &&  43.2 & 4.32 & \nodata \\ 
\ion{V}{1}  && 6090.21\tablenotemark{c}  && 1.08 && -0.062 &&  32.9 & 3.86 & 3.83 &&  66.3 & 4.27 & 4.22 &&  69.1 & 4.29 & 4.22 \\ 
\ion{V}{1}  && 6111.65\tablenotemark{c}  && 1.04 && -0.715 &&  10.7 & 3.86 & 3.81 &&  37.5 & 4.39 & 4.26 &&  41.3 & 4.41 & 4.29 \\ 
\ion{V}{1}  && 6251.83 && 0.29 && -1.340 &&  17.9 & 3.98 & \nodata &&  46.4 & 4.37 & \nodata &&  53.5 & 4.44 & \nodata \\ 
\ion{Cr}{1} && 5702.31 && 3.45 && -0.667 &&  23.1 & 5.78 & \nodata &&  49.1 & 6.19 & \nodata &&  53.4 & 6.25 & \nodata \\ 
\ion{Cr}{1} && 5783.06 && 3.32 && -0.500 &&  35.9 & 5.76 & \nodata &&  55.2 & 6.01 & \nodata && \nodata & \nodata & \nodata \\ 
\ion{Cr}{1} && 5783.85 && 3.32 && -0.295 &&  43.0 & 5.69 & \nodata &&  72.0 & 6.08 & \nodata &&  73.3 & 6.09 & \nodata \\ 
\ion{Cr}{1} && 5787.92 && 3.32 && -0.083 &&  44.8 & 5.51 & \nodata &&  69.8 & 5.83 & \nodata &&  70.5 & 5.83 & \nodata \\ 
\ion{Cr}{1} && 7400.25 && 2.90 && -0.111 &&  77.1 & 5.58 & \nodata && 107.6 & 5.93 & \nodata && 109.2 & 5.93 & \nodata \\ 
\ion{Mn}{1} && 5432.55\tablenotemark{c} && 0.00 && -3.795 && 46.1 & 5.38 & 5.27 &&  43.5 & 5.37 & 5.21 && 111.1 & 6.25 & 5.66 \\ 
\ion{Fe}{1} && 5322.04 && 2.28 && -2.800 &&  57.6 & 7.20 & \nodata &&  80.6 & 7.52 & \nodata &&  81.0 & 7.54 & \nodata \\ 
\ion{Fe}{1} && 5379.57 && 3.69 && -1.510 &&  58.4 & 7.30 & \nodata &&  84.6 & 7.69 & \nodata &&  81.6 & 7.64 & \nodata \\ 
\ion{Fe}{1} && 5522.45 && 4.21 && -1.550 &&  42.7 & 7.54 & \nodata &&  61.7 & 7.81 & \nodata &&  64.4 & 7.86 & \nodata \\ 
\ion{Fe}{1} && 5543.94 && 4.22 && -1.140 &&  59.5 & 7.43 & \nodata &&  82.9 & 7.77 & \nodata &&  85.4 & 7.81 & \nodata \\ 
\ion{Fe}{1} && 5546.50 && 4.37 && -1.310 &&  49.7 & 7.57 & \nodata &&  71.4 & 7.88 & \nodata &&  73.0 & 7.91 & \nodata \\ 
\ion{Fe}{1} && 5546.99 && 4.22 && -1.910 &&  27.9 & 7.61 & \nodata &&  55.7 & 8.07 & \nodata &&  57.7 & 8.10 & \nodata \\ 
\ion{Fe}{1} && 5560.21 && 4.43 && -1.190 &&  49.6 & 7.50 & \nodata &&  69.3 & 7.79 & \nodata &&  69.3 & 7.78 & \nodata \\ 
\ion{Fe}{1} && 5577.03 && 5.03 && -1.550 &&  12.0 & 7.55 & \nodata &&  23.3 & 7.85 & \nodata &&  23.8 & 7.85 & \nodata \\ 
\ion{Fe}{1} && 5579.34 && 4.23 && -2.400 &&  11.1 & 7.61 & \nodata &&  24.5 & 7.96 & \nodata &&  24.1 & 7.94 & \nodata \\ 
\ion{Fe}{1} && 5587.57 && 4.14 && -1.850 &&  37.6 & 7.67 & \nodata &&  59.5 & 8.00 & \nodata &&  59.5 & 7.99 & \nodata \\ 
\ion{Fe}{1} && 5651.47 && 4.47 && -2.000 &&  18.5 & 7.70 & \nodata &&  34.9 & 8.02 & \nodata &&  35.5 & 8.02 & \nodata \\ 
\ion{Fe}{1} && 5652.32 && 4.26 && -1.950 &&  25.8 & 7.64 & \nodata &&  45.0 & 7.96 & \nodata &&  46.0 & 7.97 & \nodata \\ 
\ion{Fe}{1} && 5661.35 && 4.28 && -1.740 &&  21.6 & 7.34 & \nodata &&  41.5 & 7.70 & \nodata &&  42.4 & 7.71 & \nodata \\ 
\ion{Fe}{1} && 5667.52 && 4.18 && -1.580 &&  49.2 & 7.65 & \nodata &&  78.6 & 8.09 & \nodata &&  79.1 & 8.10 & \nodata \\ 
\ion{Fe}{1} && 5677.68 && 4.10 && -2.700 &&   6.5 & 7.52 & \nodata &&  15.8 & 7.89 & \nodata &&  16.6 & 7.90 & \nodata \\ 
\ion{Fe}{1} && 5679.02 && 4.65 && -0.920 &&  57.7 & 7.56 & \nodata &&  77.6 & 7.83 & \nodata &&  78.8 & 7.85 & \nodata \\ 
\ion{Fe}{1} && 5680.24 && 4.19 && -2.580 &&   9.8 & 7.68 & \nodata &&  22.7 & 8.05 & \nodata &&  24.0 & 8.07 & \nodata \\ 
\ion{Fe}{1} && 5732.27 && 4.99 && -1.560 &&  13.8 & 7.59 & \nodata &&  29.3 & 7.96 & \nodata &&  29.8 & 7.96 & \nodata \\ 
\ion{Fe}{1} && 5741.85 && 4.26 && -1.850 &&  31.0 & 7.65 & \nodata &&  51.8 & 7.97 & \nodata &&  52.9 & 7.98 & \nodata \\ 
\ion{Fe}{1} && 5752.03 && 4.55 && -1.180 &&  54.1 & 7.67 & \nodata &&  73.6 & 7.94 & \nodata &&  73.3 & 7.94 & \nodata \\ 
\ion{Fe}{1} && 5775.08 && 4.22 && -1.300 &&  55.8 & 7.52 & \nodata &&  79.4 & 7.86 & \nodata &&  79.6 & 7.86 & \nodata \\ 
\ion{Fe}{1} && 5778.45 && 2.59 && -3.480 &&  21.7 & 7.44 & \nodata &&  43.9 & 7.79 & \nodata &&  44.5 & 7.78 & \nodata \\ 
\ion{Fe}{1} && 6079.00 && 4.65 && -1.120 &&  43.8 & 7.52 & \nodata &&  64.3 & 7.81 & \nodata &&  66.7 & 7.85 & \nodata \\ 
\ion{Fe}{1} && 6085.26 && 2.76 && -3.100 &&  42.2 & 7.64 & \nodata &&  71.3 & 8.05 & \nodata &&  75.3 & 8.13 & \nodata \\ 
\ion{Fe}{1} && 6098.24 && 4.56 && -1.880 &&  16.4 & 7.58 & \nodata &&  36.0 & 7.99 & \nodata &&  33.2 & 7.92 & \nodata \\ 
\ion{Fe}{1} && 6151.62 && 2.18 && -3.300 &&  48.1 & 7.37 & \nodata &&  70.6 & 7.64 & \nodata &&  72.5 & 7.67 & \nodata \\ 
\ion{Fe}{1} && 6159.37 && 4.61 && -1.970 &&  12.3 & 7.57 & \nodata &&  27.3 & 7.95 & \nodata &&  29.2 & 7.98 & \nodata \\ 
\ion{Fe}{1} && 6165.36 && 4.14 && -1.470 &&  43.6 & 7.38 & \nodata &&  65.5 & 7.69 & \nodata &&  65.4 & 7.69 & \nodata \\ 
\ion{Fe}{1} && 6187.99 && 3.94 && -1.720 &&  46.2 & 7.48 & \nodata &&  71.6 & 7.85 & \nodata &&  73.3 & 7.88 & \nodata \\ 
\ion{Fe}{1} && 6220.78 && 3.88 && -2.460 &&  18.6 & 7.57 & \nodata &&  37.3 & 7.92 & \nodata &&  39.8 & 7.96 & \nodata \\ 
\ion{Fe}{1} && 6226.73 && 3.88 && -2.220 &&  28.3 & 7.57 & \nodata &&  54.7 & 8.00 & \nodata &&  51.2 & 7.93 & \nodata \\ 
\ion{Fe}{1} && 6229.23 && 2.85 && -2.810 &&  37.7 & 7.35 & \nodata &&  64.1 & 7.72 & \nodata &&  62.4 & 7.68 & \nodata \\ 
\ion{Fe}{1} && 6240.65 && 2.22 && -3.230 &&  47.7 & 7.32 & \nodata &&  75.8 & 7.69 & \nodata &&  75.2 & 7.68 & \nodata \\ 
\ion{Fe}{1} && 6380.74 && 4.19 && -1.380 &&  52.2 & 7.48 & \nodata &&  77.1 & 7.83 & \nodata &&  72.6 & 7.76 & \nodata \\ 
\ion{Fe}{1} && 6608.02 && 2.28 && -4.030 &&  17.1 & 7.50 & \nodata &&  37.4 & 7.85 & \nodata &&  39.2 & 7.86 & \nodata \\ 
\ion{Fe}{1} && 6609.11 && 2.56 && -2.690 &&  65.1 & 7.40 & \nodata &&  90.9 & 7.74 & \nodata &&  92.1 & 7.76 & \nodata \\ 
\ion{Fe}{1} && 6627.54 && 4.55 && -1.680 &&  29.4 & 7.68 & \nodata &&  52.3 & 8.05 & \nodata &&  50.3 & 8.01 & \nodata \\ 
\ion{Fe}{1} && 6653.85 && 4.15 && -2.520 &&  10.9 & 7.60 & \nodata &&  21.3 & 7.88 & \nodata &&  24.1 & 7.93 & \nodata \\ 
\ion{Fe}{1} && 6703.57 && 2.76 && -3.160 &&  36.2 & 7.55 & \nodata &&  62.4 & 7.90 & \nodata &&  60.1 & 7.85 & \nodata \\ 
\ion{Fe}{1} && 6710.32 && 1.49 && -4.880 &&  18.8 & 7.61 & \nodata &&  42.1 & 7.96 & \nodata &&  40.6 & 7.90 & \nodata \\ 
\ion{Fe}{1} && 6713.74 && 4.80 && -1.600 &&  20.2 & 7.61 & \nodata &&  40.5 & 8.00 & \nodata &&  39.2 & 7.97 & \nodata \\ 
\ion{Fe}{1} && 6716.22 && 4.58 && -1.920 &&  15.5 & 7.58 & \nodata &&  31.4 & 7.93 & \nodata &&  30.9 & 7.91 & \nodata \\ 
\ion{Fe}{1} && 6725.35 && 4.10 && -2.300 &&  17.4 & 7.56 & \nodata &&  34.7 & 7.90 & \nodata &&  36.5 & 7.92 & \nodata \\ 
\ion{Fe}{1} && 6726.67 && 4.61 && -1.130 &&  45.8 & 7.49 & \nodata &&  71.0 & 7.86 & \nodata &&  69.3 & 7.83 & \nodata \\ 
\ion{Fe}{1} && 6733.15 && 4.64 && -1.580 &&  26.4 & 7.59 & \nodata &&  49.4 & 7.98 & \nodata &&  48.4 & 7.95 & \nodata \\ 
\ion{Fe}{1} && 6739.52 && 1.56 && -4.790 &&  11.8 & 7.35 & \nodata &&  27.8 & 7.67 & \nodata &&  30.3 & 7.69 & \nodata \\ 
\ion{Fe}{1} && 6745.09 && 4.58 && -2.160 &&   8.6 & 7.53 & \nodata &&  19.0 & 7.87 & \nodata &&  19.2 & 7.86 & \nodata \\ 
\ion{Fe}{1} && 6745.96 && 4.08 && -2.770 &&   7.6 & 7.60 & \nodata &&  16.1 & 7.90 & \nodata &&  17.4 & 7.93 & \nodata \\ 
\ion{Fe}{1} && 6752.72 && 4.64 && -1.300 &&  34.6 & 7.49 & \nodata &&  59.1 & 7.86 & \nodata &&  58.3 & 7.85 & \nodata \\ 
\ion{Fe}{2} && 5197.58 && 3.23 && -2.348 &&  77.1 & 7.32 & \nodata &&  86.7 & 7.69 & \nodata &&  80.0 & 7.63 & \nodata \\ 
\ion{Fe}{2} && 5234.62 && 3.22 && -2.279 &&  77.8 & 7.26 & \nodata &&  88.5 & 7.64 & \nodata &&  86.9 & 7.69 & \nodata \\ 
\ion{Fe}{2} && 5414.07 && 3.22 && -3.645 &&  28.6 & 7.56 & \nodata &&  32.4 & 7.80 & \nodata &&  31.7 & 7.84 & \nodata \\ 
\ion{Fe}{2} && 6084.11 && 3.20 && -3.881 &&  20.0 & 7.53 & \nodata &&  27.3 & 7.89 & \nodata &&  25.6 & 7.89 & \nodata \\ 
\ion{Fe}{2} && 6113.32 && 3.22 && -4.230 &&  12.5 & 7.64 & \nodata &&  16.8 & 7.94 & \nodata &&  17.6 & 8.01 & \nodata \\ 
\ion{Fe}{2} && 6149.26 && 3.89 && -2.841 &&  34.5 & 7.51 & \nodata &&  45.6 & 7.93 & \nodata &&  39.1 & 7.85 & \nodata \\ 
\ion{Fe}{2} && 6239.95 && 3.89 && -3.573 &&  13.3 & 7.63 & \nodata &&  19.3 & 8.00 & \nodata &&  17.8 & 8.00 & \nodata \\ 
\ion{Fe}{2} && 6247.56 && 3.89 && -2.435 &&  51.2 & 7.46 & \nodata &&  58.5 & 7.80 & \nodata &&  54.5 & 7.79 & \nodata \\ 
\ion{Co}{1} && 5301.04\tablenotemark{c} && 1.71 && -2.000 &&  19.9 & 4.96 & \nodata &&  5.9 & 4.25 & 5.40 &&  50.4 & 5.47 & 5.40 \\ 
\ion{Co}{1} && 5647.23 && 2.28 && -1.560 &&  14.9 & 4.90 & \nodata &&  35.7 & 5.32 & \nodata &&  37.8 & 5.35 & \nodata \\ 
\ion{Co}{1} && 6093.14 && 1.74 && -2.440 &&   8.5 & 4.94 & \nodata &&  26.7 & 5.43 & \nodata && \nodata & \nodata & \nodata \\ 
\ion{Co}{1} && 6678.80 && 1.96 && -2.680 &&   6.3 & 5.22 & \nodata &&  17.7 & 5.63 & \nodata &&  20.5 & 5.68 & \nodata \\ 
\ion{Co}{1} && 6814.94\tablenotemark{c} && 1.96 && -1.900 &&  18.8 & 4.97 & \nodata &&  49.5 & 5.50 & 5.42 &&  51.5 & 5.52 & 5.46 \\ 
\ion{Ni}{1} && 5748.35 && 1.68 && -3.260 &&  27.7 & 6.18 & \nodata &&  54.4 & 6.61 & \nodata &&  55.7 & 6.63 & \nodata \\ 
\ion{Ni}{1} && 5754.65 && 1.94 && -2.330 &&  75.9 & 6.43 & \nodata && 105.4 & 6.90 & \nodata && 107.3 & 6.95 & \nodata \\ 
\ion{Ni}{1} && 5760.83 && 4.11 && -0.800 &&  34.1 & 6.30 & \nodata &&  55.6 & 6.59 & \nodata &&  54.5 & 6.58 & \nodata \\ 
\ion{Ni}{1} && 5846.99 && 1.68 && -3.210 &&  22.2 & 6.00 & \nodata &&  47.6 & 6.43 & \nodata &&  48.7 & 6.44 & \nodata \\ 
\ion{Ni}{1} && 6111.07 && 4.09 && -0.870 &&  34.0 & 6.26 & \nodata &&  60.2 & 6.72 & \nodata &&  59.6 & 6.71 & \nodata \\ 
\ion{Ni}{1} && 6128.96 && 1.68 && -3.330 &&  25.8 & 6.19 & \nodata &&  50.0 & 6.57 & \nodata &&  51.0 & 6.58 & \nodata \\ 
\ion{Ni}{1} && 6130.13 && 4.27 && -0.960 &&  21.4 & 6.23 & \nodata &&  40.0 & 6.61 & \nodata &&  40.3 & 6.62 & \nodata \\ 
\ion{Ni}{1} && 6133.96 && 4.09 && -1.830 &&   4.9 & 6.20 & \nodata &&  13.4 & 6.65 & \nodata &&  14.2 & 6.68 & \nodata \\ 
\ion{Ni}{1} && 6175.36 && 4.09 && -0.559 &&  47.6 & 6.20 & \nodata &&  68.6 & 6.55 & \nodata &&  70.2 & 6.59 & \nodata \\ 
\ion{Ni}{1} && 6176.81 && 4.09 && -0.260 &&  61.8 & 6.15 & \nodata &&  86.4 & 6.54 & \nodata &&  89.1 & 6.59 & \nodata \\ 
\ion{Ni}{1} && 6177.24 && 1.83 && -3.500 &&  15.4 & 6.22 & \nodata &&  33.9 & 6.60 & \nodata &&  36.0 & 6.63 & \nodata \\ 
\ion{Ni}{1} && 6186.71 && 4.11 && -0.960 &&  31.1 & 6.30 & \nodata &&  52.0 & 6.67 & \nodata &&  52.8 & 6.69 & \nodata \\ 
\ion{Ni}{1} && 6223.98 && 4.11 && -0.910 &&  28.4 & 6.20 & \nodata &&  51.4 & 6.61 & \nodata &&  50.2 & 6.60 & \nodata \\ 
\ion{Ni}{1} && 6230.09 && 4.11 && -1.260 &&  22.4 & 6.40 & \nodata &&  42.4 & 6.80 & \nodata &&  44.1 & 6.83 & \nodata \\ 
\ion{Ni}{1} && 6327.59 && 1.68 && -3.150 &&  40.0 & 6.28 & \nodata &&  65.4 & 6.65 & \nodata &&  70.1 & 6.74 & \nodata \\ 
\ion{Ni}{1} && 6370.34 && 3.54 && -1.940 &&  12.7 & 6.23 & \nodata &&  30.7 & 6.69 & \nodata &&  30.7 & 6.69 & \nodata \\ 
\ion{Ni}{1} && 6378.25 && 4.15 && -0.830 &&  34.5 & 6.28 & \nodata &&  55.0 & 6.63 & \nodata &&  57.2 & 6.68 & \nodata \\ 
\ion{Ni}{1} && 6643.63 && 1.68 && -2.300 &&  93.6 & 6.37 & \nodata && 124.0 & 6.78 & \nodata && 128.2 & 6.86 & \nodata \\ 
\ion{Ni}{1} && 6767.77 && 1.83 && -2.170 &&  76.5 & 6.08 & \nodata && 104.8 & 6.48 & \nodata && 106.4 & 6.53 & \nodata \\ 
\enddata
\tablenotetext{a}{Indicates the $\log{N}$ abundance determined from the synthetic fit to a given line. Each synthetic
                  fit was performed with the {\sc moog} {\it synth} driver. Synthetic fits were only 
                  performed for the subset of V, Mn, and Co lines that were tested for hfs.}
\tablenotetext{c}{Indicates that the spectral line was tested for hfs.}
\end{deluxetable}


\begin{thebibliography}{50}
\expandafter\ifx\csname natexlab\endcsname\relax\def\natexlab#1{#1}\fi
\bibitem[Adibekyan et 
al.(2014)]{2014A&A...564L..15A} Adibekyan, V.~Z., Gonz{\'a}lez Hern{\'a}ndez, J.~I., Delgado Mena, E., et al.\ 2014, \aap, 564, L15

\bibitem[{{Asplund} {et~al.}(2009){Asplund}, {Grevesse}, {Sauval}, \&
  {Scott}}]{2009ARA&A..47..481A}
{Asplund}, M., {Grevesse}, N., {Sauval}, A.~J., \& {Scott}, P. 2009, \araa, 47,
  481
  
\bibitem[Biazzo et al.(2015)]{biazzo15} Biazzo, K., Gratton, R., 
Desidera, S., et al.\ 2015, arXiv:1506.01614 

\bibitem[Campante et al.(2015)]{campante15} Campante, T.~L., 
Barclay, T., Swift, J.~J., et al.\ 2015, \apj, 799, 170 

\bibitem[Chambers(2010)]{2010ApJ...724...92C} Chambers, J.~E.\ 2010, \apj, 
724, 92 

\bibitem[Deal et al.(2015)]{deal15} Deal, M., Richard, O., 
\& Vauclair, S.\ 2015, arXiv:1509.06958 

\bibitem[{{ESA}(1997)}]{1997ESASP1200.....E}
{ESA}, ed. 1997, ESA Special Publication, Vol. 1200, {The HIPPARCOS and TYCHO
  catalogues. Astrometric and photometric star catalogues derived from the ESA
  HIPPARCOS Space Astrometry Mission}

\bibitem[{{Fitzpatrick} \& {Sneden}(1987)}]{1987BAAS...19.1129F}
{Fitzpatrick}, M.~J., \& {Sneden}, C. 1987, in Bulletin of the American
  Astronomical Society, Vol.~19, Bulletin of the American Astronomical Society,
  1129

\bibitem[{{Johnson} {et~al.}(2006){Johnson}, {Ivans}, \&
  {Stetson}}]{2006ApJ...640..801J}
{Johnson}, J.~A., {Ivans}, I.~I., \& {Stetson}, P.~B. 2006, \apj, 640, 801

\bibitem[{{Kharchenko}(2001)}]{2001KFNT...17..409K}
{Kharchenko}, N.~V. 2001, Kinematika i Fizika Nebesnykh Tel, 17, 409

\bibitem[{{Kupka} {et~al.}(1999){Kupka}, {Piskunov}, {Ryabchikova}, {Stempels},
  \& {Weiss}}]{1999A&AS..138..119K}
{Kupka}, F., {Piskunov}, N., {Ryabchikova}, T.~A., {Stempels}, H.~C., \&
  {Weiss}, W.~W. 1999, \aaps, 138, 119
  
\bibitem[Liu et al.(2014)]{lui14} Liu, F., Asplund, M., 
Ram{\'{\i}}rez, I., Yong, D., \& Mel{\'e}ndez, J.\ 2014, \mnras, 442, L51 

\bibitem[{{Lodders}(2003)}]{2003ApJ...591.1220L}
{Lodders}, K. 2003, \apj, 591, 1220

\bibitem[{{Mack} {et~al.}(2014){Mack}, {Schuler}, {Stassun}, \&
  {Norris}}]{mack14}
{Mack}, III, C.~E., {Schuler}, S.~C., {Stassun}, K.~G., \& {Norris}, J. 2014,
  \apj, 787, 98

\bibitem[Metcalfe et al.(2012)]{metcalfe12} Metcalfe, T.~S., 
Chaplin, W.~J., Appourchaux, T., et al.\ 2012, \apjl, 748, L10 

\bibitem[Metcalfe et al.(2015)]{metcalfe15} Metcalfe, T.~S., 
Creevey, O.~L., \& Davies, G.~R.\ 2015, \apjl, 811, L37 

\bibitem[Mel{\'e}ndez et al.(2009)]{2009ApJ...704L..66M} Mel{\'e}ndez, J., 
Asplund, M., Gustafsson, B., \& Yong, D.\ 2009, \apjl, 704, L66 

\bibitem[McDonough(2001)]{McDonough01} {McDonough}, W. 2001, in The Composition of the Earth,
in Earthquake Thermodynamics and Phase Transitions in the Earth's Interior (International Geophysics
Series, Vol. 76), ed. R. Teisseyre \& E. Majewski (San Diego, CA: Academic Press)

\bibitem[Naef et 
al.(2001)]{naef01} Naef, D., Latham, D.~W., Mayor, M., et al.\ 2001, \aap, 375, L27 


\bibitem[{{Pinsonneault} {et~al.}(2001){Pinsonneault}, {DePoy}, \&
  {Coffee}}]{2001ApJ...556L..59P}
{Pinsonneault}, M.~H., {DePoy}, D.~L., \& {Coffee}, M. 2001, \apjl, 556, L59

\bibitem[{{Piskunov} {et~al.}(1995){Piskunov}, {Kupka}, {Ryabchikova}, {Weiss},
  \& {Jeffery}}]{1995A&AS..112..525P}
{Piskunov}, N.~E., {Kupka}, F., {Ryabchikova}, T.~A., {Weiss}, W.~W., \&
  {Jeffery}, C.~S. 1995, \aaps, 112, 525

\bibitem[{{Pont} {et~al.}(2009){Pont}, {H{\'e}brard}, {Irwin}, {Bouchy},
  {Moutou}, {Ehrenreich}, {Guillot}, {Aigrain}, {Bonfils}, {Berta}, {Boisse},
  {Burke}, {Charbonneau}, {Delfosse}, {Desort}, {Eggenberger}, {Forveille},
  {Lagrange}, {Lovis}, {Nutzman}, {Pepe}, {Perrier}, {Queloz}, {Santos},
  {S{\'e}gransan}, {Udry}, \& {Vidal-Madjar}}]{2009A&A...502..695P}
{Pont}, F., {H{\'e}brard}, G., {Irwin}, J.~M., {et~al.} 2009, \aap, 502, 695

\bibitem[{{Prochaska} \& {McWilliam}(2000)}]{2000ApJ...537L..57P}
{Prochaska}, J.~X., \& {McWilliam}, A. 2000, \apjl, 537, L57

\bibitem[Ram{\'{\i}}rez et al.(2015)]{2015ApJ...808...13R} Ram{\'{\i}}rez, 
I., Khanal, S., Aleo, P., et al.\ 2015, \apj, 808, 13

\bibitem[Ram{\'{\i}}rez et 
al.(2009)]{2009A&A...508L..17R} Ram{\'{\i}}rez, I., Mel{\'e}ndez, J., \& Asplund, M.\ 2009, \aap, 508, L17

\bibitem[Ram{\'{\i}}rez et al.(2011)]{ramirez11} Ram{\'{\i}}rez, 
I., Mel{\'e}ndez, J., Cornejo, D., Roederer, I.~U., 
\& Fish, J.~R.\ 2011, \apj, 740, 76 

\bibitem[Saffe et 
al.(2015)]{saffe15} Saffe, C., Flores, M., \& Buccino, A.\ 2015, \aap, 582, A17 

\bibitem[{{Schuler} {et~al.}(2011a){Schuler}, {Flateau}, {Cunha}, {King},
  {Ghezzi}, \& {Smith}}]{schuler11a}
{Schuler}, S.~C., {Flateau}, D., {Cunha}, K., {et~al.} 2011a, \apj, 732, 55

\bibitem[Schuler et al.(2011b)]{schuler11b} Schuler, S.~C., Cunha, 
K., Smith, V.~V., et al.\ 2011b, \apjl, 737, L32 

\bibitem[Schuler et al.(2015)]{2015ApJ...815....5S} Schuler, S.~C., Vaz, 
Z.~A., Katime Santrich, O.~J., et al.\ 2015, \apj, 815, 5

\bibitem[{{Sneden}(1973)}]{1973PhDT.......180S}
{Sneden}, C.~A. 1973, PhD thesis, THE UNIVERSITY OF TEXAS AT AUSTIN.

\bibitem[Teske et al.(2015)]{teske15} Teske, J.~K., Ghezzi, L., 
Cunha, K., et al.\ 2015, \apjl, 801, L10 

\bibitem[{{Torres} {et~al.}(2010){Torres}, {Andersen}, \&
  {Gim{\'e}nez}}]{2010A&ARv..18...67T}
{Torres}, G., {Andersen}, J., \& {Gim{\'e}nez}, A. 2010, \aapr, 18, 67

\bibitem[Tucci Maia et al.(2014)]{tucci14} Tucci Maia, M., 
Mel{\'e}ndez, J., \& Ram{\'{\i}}rez, I.\ 2014, \apjl, 790, L25 

\bibitem[{{Vogt} {et~al.}(1994){Vogt}, {Allen}, {Bigelow}, {Bresee}, {Brown},
  {Cantrall}, {Conrad}, {Couture}, {Delaney}, {Epps}, {Hilyard}, {Hilyard},
  {Horn}, {Jern}, {Kanto}, {Keane}, {Kibrick}, {Lewis}, {Osborne},
  {Pardeilhan}, {Pfister}, {Ricketts}, {Robinson}, {Stover}, {Tucker}, {Ward},
  \& {Wei}}]{1994SPIE.2198..362V}
{Vogt}, S.~S., {Allen}, S.~L., {Bigelow}, B.~C., {et~al.} 1994, in Society of
  Photo-Optical Instrumentation Engineers (SPIE) Conference Series, Vol. 2198,
  Instrumentation in Astronomy VIII, ed. D.~L. {Crawford} \& E.~R. {Craine},
  362
  
\end{thebibliography}
\end{document}